\documentclass[review]{elsarticle}
\usepackage{lineno,hyperref}
\usepackage{amsmath}
\usepackage{subfig}
\usepackage{bm}
\usepackage{natbib}
\modulolinenumbers[5]

\journal{ArXiv}

\begin{document}

\begin{frontmatter}

\title{Evolution of interfacial dislocation network during
low-stress high temperature creep of particle strengthened alloy
system using discrete dislocation dynamics}

%\author{\fnref{myfootnote}}

\author[mymainaddress]{Tushar Jogi}
\ead{ms14resch11003@iith.ac.in}

\author[mymainaddress]{Saswata Bhattacharya
\corref{mycorrespondingauthor}}
\cortext[mycorrespondingauthor]{Corresponding author}
\ead{saswata@iith.ac.in}

\address[mymainaddress]
{Department of Materials Science and Metallurgical	Engineering, 
Indian Institute of Technology Hyderabad, 
Kandi, Sangareddy-502285}

\begin{abstract}
We use three-dimensional discrete dislocation dynamics (DDD)
simulations to study the evolution of interfacial dislocation network (IDN) 
in particle-strengthened alloy systems subjected to a constant stress 
at high temperatures. We have modified the dislocation mobility laws 
to incorporate the recovery of dislocation network by climb.  
The microstructure consists of uniformly distributed cuboidal  
inclusions embedded in the simulation box. 
Based on the systematic simulations of IDN formation as a function of 
applied stress for prescribed inter-particle spacing and  glide-to-climb 
mobility ratio,  we derive a relation  between effective stress and 
normalized dislocation density. We use link-length analysis to show self 
similarity of immobile dislocation links irrespective of the level of applied 
stress. Moreover, we modify Taylor's relation to show the dependence of 
effective stress on ratio between mobile to immobile dislocation density. We 
justify the relation with the help of a theoretical model which takes into 
account the balance of multiplication and annihilation rates of dislocation 
density. 

\end{abstract}

\begin{keyword}
discrete dislocation dynamics, creep, superalloy
\end{keyword}

\end{frontmatter}

%\linenumbers

The dislocation movement and their interactions with underlying microstructure
decide the life of $\gamma^{\prime} $ strengthened Ni-base superalloys. 
These alloys exhibit exceptional mechanical strength at high temperature 
because of precipitation strengthening where coherent $\gamma^ {\prime}$ 
precipitates impedes the motion of dislocation through matrix
~\cite{reed2008superalloys}. 
Many TEM observations have revealed the formation
of dense interfacial dislocation network (IDN) around cuboidal 
$\gamma^{\prime}$ precipitates during low to intermediate stresses 
($100-550MPa$) and high homologous temperature ($850-1050^\circ C$) creep 
deformation of these materials~\cite{Lasalmonie1975,Feller-Kniepmeier1989,
Pollock1992,Nategh2003,carroll2008interfacial,ru2016dislocation}. 
Several groups have reported that the evolution of the dense network of
dislocation govern the recovery process during creep. 
Over the years, researchers have developed sophisticated electron microscopy 
techniques to investigate the formation of dislocation networks high 
temperature creep~\cite{Gabb1989,Sugui2000,Zhang2002,Zhang2003,Zhang2005,
Epishin2007}. 
Most of these investigations show the IDN over the surface of 
precipitates. 
They have proposed that such network formed as a result of 
dislocation reactions. 
A steady state high temperature creep evolution
in particle strengthened alloy system can be visualized as development of
three dimensional IDN network and consecutive occurrence of strain hardening 
and recovery events~\cite{Lagneborg1968, Lagneborg1969, Lagneborg1969a}.
Attractive junctions formed due to dislocation reactions contribute to strain
hardening.
Under the influence of applied stress, eventually the some of the
attractive junctions break down and released dislocation gives rise to 
recovery event. Further, the released dislocation again conjoins the network 
and consecutive strain hardening and recovery events are repeated. 

\par
Most studies conclude that the characteristics of the 
dislocation network is a complex byproduct of applied stress, lattice misfit, 
and interparticle spacing~\cite{Gabb1989,yue2018stress}.
Zhang et.al.~\cite{Zhang2003} have 
studied the role of sign and magnitude of lattice mismatch between $\gamma$ 
and coherent $\gamma^{\prime}$ precipitates on the mobile characteristics of 
the network. 
Larger negative misfit transforms more mobile hexagonal morphology to less 
mobile square network. In some cases, dislocation network contains 
a mixture of square and hexagonal morphology~\cite{ru2016dislocation,
field1992development}. 
Several studies based on computer simulations in three dimensions have been 
conducted to understand the mechanisms of formation of IDN and its role 
during the high temperature creep at continuum as well as atomic length 
scales.
Yashiro et. al.~\cite{Yashiro2008} investigated the stability of interfacial 
dislocation network and introduced a softening mechanisms of dislocation 
penetration through precipitate by using back-force model in DDD framework. 
Further, using back-force model, Huang et. al.~\cite{Huang2012} investigated 
the effect of precipitate size, channel width, and shape of precipitate 
(spherical and cuboidal) on formation of dislocation network and hardening 
rate. 
In another work, Haghighat et. al.~\cite{HafezHaghighat2013} reported 
a DDD model wherein the dislocation-precipitate interaction forces and the 
conservative climb of edge dislocation as recovery mechanism during high 
temperature creep is taken into account. 
Furthermore, Liu et. al. 
~\cite{Liu2014} extended the model by incorporating misfit stresses to 
understand formation of different dislocation junctions (e.g. Lomer junction, 
Hirth junction, glissile junction, etc.) arising due to dislocation reactions 
and studied its effect on dynamic recovery during high temperature creep. 
Using DDD simulations, Gao et. al.~\cite{gao2015influence} have investigated 
the effect of sign of lattice misfit and sign of applied stress (compressive 
or tensile) on deformation mechanisms and anisotropy of deformation in 
$\gamma$ channels.
\par 
Zhu et. al.~\cite{zhu2013atomistic} studied the formation of 
interfacial dislocation network using molecular dynamics simulations where 
stability of dislocation network was shown to be due to the formation of
Lomer-Cottrel junctions and dislocation network provides resistance to matrix 
dislocation to penetrate the particles. 
\par
Many studies have analysed the dislocation network using link-length
distribution model where the statistical distribution of dislocation 
link length gives the useful measures of high temperature creep
~\cite{lin1989scaling,shi1993dislocation,ardell1984dislocation,
ostrom1976recovery}.
There exists a stress dependent threshold dislocation link length in the
distribution below which all the dislocation links are sessile whereas
remaining region consists of mobile dislocation links. A steady state of 
dislocation network is realised when the dislocation link distribution is
time independent. Thus, the ratio of mobile dislocation density to sessile 
dislocation density is constant at steady state. Sills et. al.~\cite{Sills2017}
reported that dislocation link length follows an exponential distribution 
using DDD simulations of strain hardening in single crystals at room 
temperature and showed that the Taylor's relation is obeyed in DDD 
simulations. However, at high temperatures (above $0.5 T_m$) the validity
of Taylors' relation is not known with certainty~\cite{mecking1981kinetics}.
\par
In this work, we investigate the mechanisms 
for the formation of the interfacial dislocation network with the 
help of DDD method. Besides, we obtain the relationship between 
normalized dislocation density and normalized effective stress at the 
steady state high temperature creep and rationalize the relation 
theoretically.  

\par            
The constant stress DDD simulations of microstructure comprising of 
uniformly arranged coherent ordered cuboidal inclusions are 
performed using modified ParaDis code. The ParaDis code is 
large scale massively parallel code to perform single crystal 
discrete dislocation dynamics simulation in FCC and BCC metals. The 
detailed algorithm and exhaustive exposition to method is available 
in references~\cite{bulatov2006computer,Arsenlis2007}. The current
open source ParaDis code takes into account dislocation-dislocation 
elastic interaction and dislocation elastic self-interaction. The 
detailed model description and governing equations are elucidated in 
reference~\cite{jogi2016evolution}. However, we summarize the extensions in
the following equations. 
\begin{itemize}
  \item ParaDis code has been extended to take into account 
  particle-dislocation interaction forces arising due to order 
  strengthening (refer eqn \eqref{particleforce}).
  \begin{equation}
  f_{ij}^{particle} = 
  	\begin{cases}
  		\frac{1}{2} \frac{\chi_{APB}}{\|\mathbf{b}\|} \|\mathbf{l}_{ij}\|
  		\tanh\left( \frac{3(L- d_{min})}{L}\right) \mathbf{n^s} & d_{min} \leq L\\
  		0 & d_{min} > L
  	\end{cases} 
  	\label{particleforce}
  \end{equation}
  Where, $f_{ij}^{particle}$ is particle-dislocation forces on the segment 
  connecting node $i$ and $j$; $\chi_{APB}$ is APB energy; $\mathbf{l}$ is 
  dislocation segment vector; $\mathbf{b}$ is Burger's vector; $d_{min}$ is
  normal distance from dislocation node to surface of the particle; 
  $\mathbf{n^s}$ represents normal vector to the particle from node.
  %Although the dislocation-vacancy interaction 
  %forces and particle-dislocation interaction due to misfit 
  %strengthening and modulus mismatch strengthening can be easily 
  %take into account in the model, we disregard them.

  \item For dislocation segments lying on $\{ 1 1 1\}$ slip planes, 
  octahedral slip and climb is considered. On the other hand, for 
  dislocation segments not lying on $\{ 1 1 1 \}$ slip planes, only 
  isotropic climb is considered
  (refer \eqref{dragtensor}). 
  Climb mobility of mixed dislocation segment is described 
  as interpolation between climb mobility of edge component of 
  dislocation segment and glide mobility of screw component of 
  the dislocation segment (see eqn \eqref{climbmobility}).
  \begin{equation}
  B_{ij}(\xi) = 
  \begin{cases}
   B^g m_i m_j  + B^c n_i n_j + B^l t_i t_j, & \forall \; \mathbf{n} \in 
   \{ 1 1 1 \}\\
   B^c \delta_{ij} + (B^g - B^l) t_i t_j, & \forall \;\mathbf{n} \notin 
   \{ 1 1 1\}
  \end{cases}
  \label{dragtensor}
  \end{equation}
  where, $B^g$, $B^c$, and $B^l$ represents glide, climb, and line components
  of drag tensor $B_{ij}(\xi)$ respectively; $n_i$ represents slip plane; 
  $t_i$ represents 
  dislocation line vector; 
  $\xi$ is tangential line vector to dislocation node; 
  $m_i = \epsilon_{ijk}n_jt_k$, here $\epsilon_{ijk}$ represents Levi-Civita 
  symbol.
  \begin{equation}
  B^c = \left[
                   (B^{ec})^2 \epsilon_{kij} b_i t_j \epsilon_{kpq} b_p t_q
                 + (B^{sc})^2 (b_i t_i)^2                     
        \right]^2 
  \label{climbmobility}       
  \end{equation}
  where, $B^{ec}$ and $B^{sc}$ are drag coefficients of edge and screw 
  dislocation segments along slip normal direction. Here, $B^{sc} = B^{sg}$ 
  and $B^{sg}$ represents the drag coefficient of screw dislocation along
  glide direction.

  \item The particle in the simulation box is constructed as 
  superellipsoidal inclusions. Previous DDD studies 
  \cite{HafezHaghighat2013,Liu2014} have employ perfect cubes as 
  inclusions, despite the fact that in reality the $\gamma^{\prime}$
  particles are cuboidal shaped with rounded corners and edges. 
  Superellipsoids are perfect geometric shapes belonging to 
  superquadric family which are similar to cubes but with
  rounded corners and edges \cite{jaklic2013segmentation}.It can be described
  by following equation
  \begin{equation}
  T(x,y,z) = {\left(\frac{x - x_0}{a}\right)}^n + {\left(\frac{y - y_0}{b}\right)}^n + {\left(\frac{z - z_0}{c}\right)}^n
  \label{superellipsoid}
  \end{equation}
  where, $a$,$b$, and $c$ represents the half edge length of cuboid, $n$ is
  the exponent which should be more than 3. Here we chose the value of $n$ as
  $4$.       
\end{itemize}
\par
The simulation box composed of eight equidistant cuboidal 
inclusion is considered to perform constant stress simulations. 
Parameters used for simulations are summarized  in table 
\ref{simulationparams}. 
\begin{center}
   \begin{table}[h]
     \begin{tabular}{| c | c |}
      \hline
      Shear modulus, $\mu$ & $45 \; GPa$\\
      \hline
      Poisson's ratio, $\nu$ & $0.37$\\
      \hline
      Simulation box size & 1 $\mu m^3$\\
      \hline
      Slip systems & $[0 1 1](1 1 \bar{1})$, $[1 0 1](\bar{1} 1 1)$\\
      \hline
      Burgers vector magnitude & $0.25 \; nm$\\
      \hline
      Particle sizes & $400 \; nm$\\
      \hline 
      Channel Spacing & $100 \; nm$\\
      \hline 
      Applied stress & $300, 400, and 500\; MPa$ along $[0 0 1]$ 
      direction \\
      \hline 
      Glide to climb mobility ratio $M_g/M_c$ & $1/1000$\\
      \hline
      Climb mobility $M_c$ & $0.01 \; Pa^{-1}s^{-1}$\\ 
      \hline
     \end{tabular}
    \caption{Parameters used for different set of simulations}
  \label{simulationparams}
  \end{table}
\end{center}    
The applied stress levels of $300$, $400$, 
and $500 \; MPa$ are taken in order to understand effect of applied 
stress. Since the DDD simulations are computationally intensive, in order to 
obtain appreciable plastic strain, we have chosen higher levels of stresses 
compared to those used in standard creep experiments. 
In all set of simulations, we choose inter-particle spacing 
as $100 nm$. The initial dislocation lines are generated by the 
cross product of plane normal of particle surface and slip plane of 
the dislocation. The anti-phase boundary (APB) energy is chosen such 
that the dislocation segments are impervious to particles. We take the APB 
energy as $150 \; mJ/m^2$.  
\par
\begin{figure}[htb]%
  \centering
  \includegraphics*[width=12.5cm]{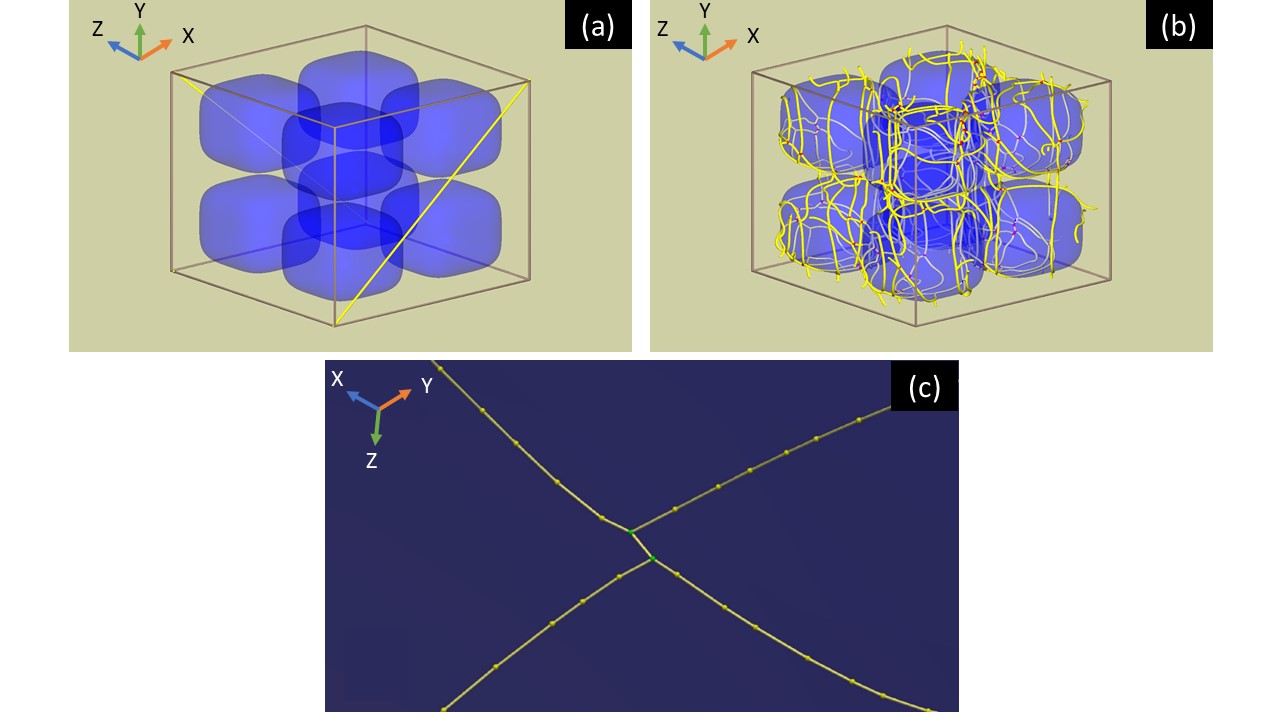}
  \caption{
    (a) Initial dislocation configuration. Two dislocation 
        line of mixed character existing on $[0 1 1](1 1 \bar{1})$ 
        and $[1 0 1](\bar{1} 1 1)$ slip systems.
    (b) The complex three interfacial dislocation network 
        formed at stress level of $300 \; MPa$. The red color nodes
        represents triple junctions.      
    (c) Formation of triple junctions by zipping at $300\;MPa$
        when two dislocation segments intersects. 
        The green color nodes represents the triple junction and segments
        held by two triple junction is newly formed sessile dislocation 
        segment.
  }
  \label{figure1}
\end{figure}

 
The initial dislocation configuration which is used for each set of
simulation is shown in fig \ref{figure1}(a). Figure \ref{figure1} 
(b) shows a three dimensional interfacial dislocation 
network formed around cuboidal particles. 
The intersection of non-coplanar dislocations produces triple junction
(triple junction is the one where three dislocation segments meet) 
(see fig \ref{figure1} (c)). The process 
is generally referred as zipping of junction. Under some critical 
stress, however, the unzipping of junction can take place wherein 
the triple junction can get destroyed. In some instances, 
since the triple junctions are more energetically favorable, even though 
quadraple junctions (quadraple junction is one where four dislocation 
segments meet) are formed, it get converted into two triple junctions. It 
identifies the fact that triple junctions are most stable which constitutes 
the complex three dimensional IDN. Every two triple junctions 
formed hold the dislocation link possessing the Burgers vector which 
is given as per energetically favorable dislocation reaction shown below     
\begin{equation}
  \frac{1}{2}[ 0 \bar{1} \bar{1}] + \frac{1}{2}[ 1 0 1] 
  \rightarrow 
  \frac{1}{2}[1 \bar{1} 0] \nonumber.
  \label {dislocreac}
\end{equation}	
Thus, the Schmid factor associated with newly formed dislocation segment 
is zero. Consequently, the immobile links 
act as a bridge between four mobile dislocation segments. This 
leads to an implication that immobile dislocation links critically 
controls dynamics of interfacial dislocation network.

\par
\begin{figure}[!htb]%
  \centering
    \subfloat[]{\includegraphics[width=7.0cm]{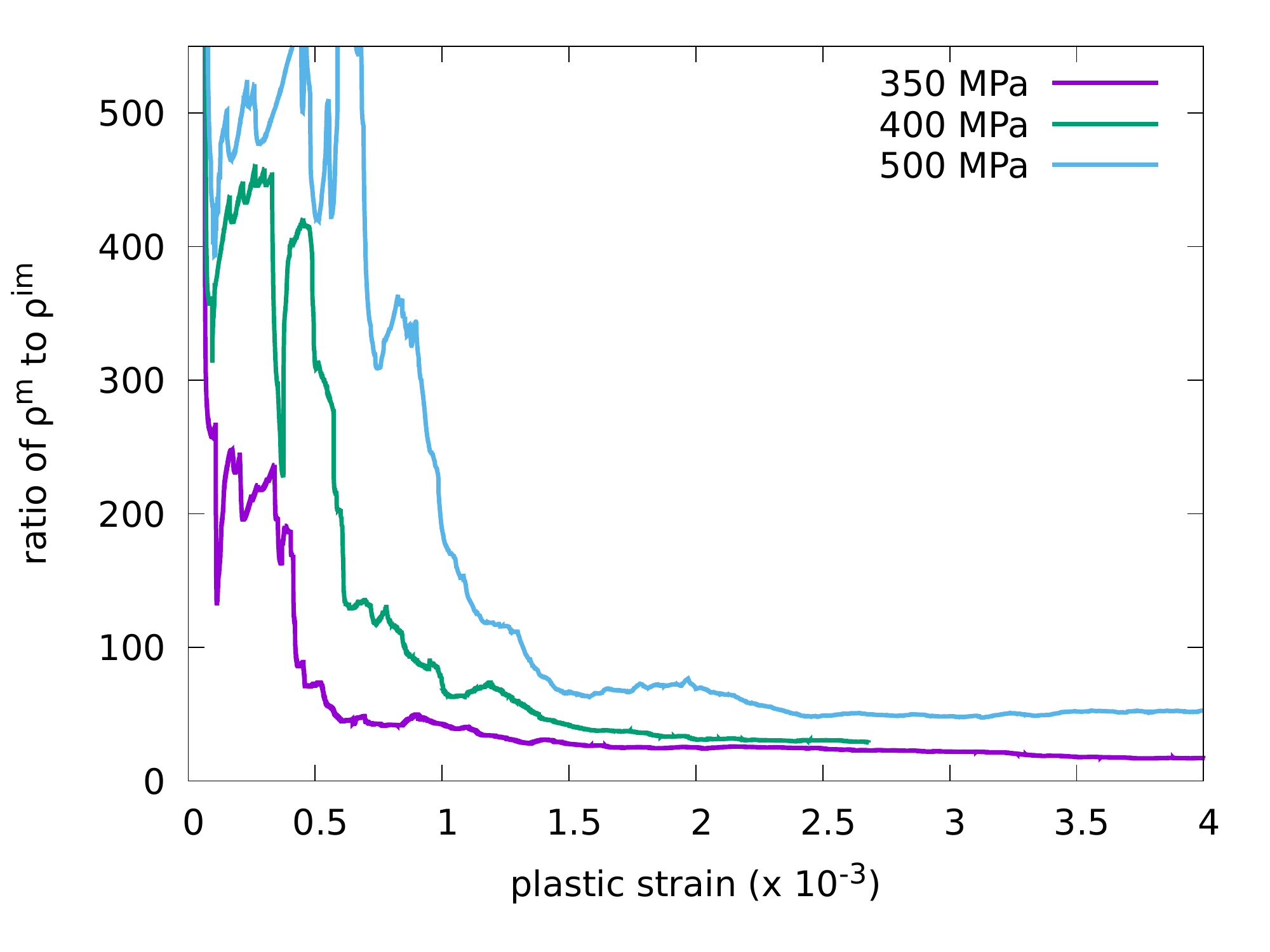}}
    \subfloat[]{\includegraphics[width=7.0cm]{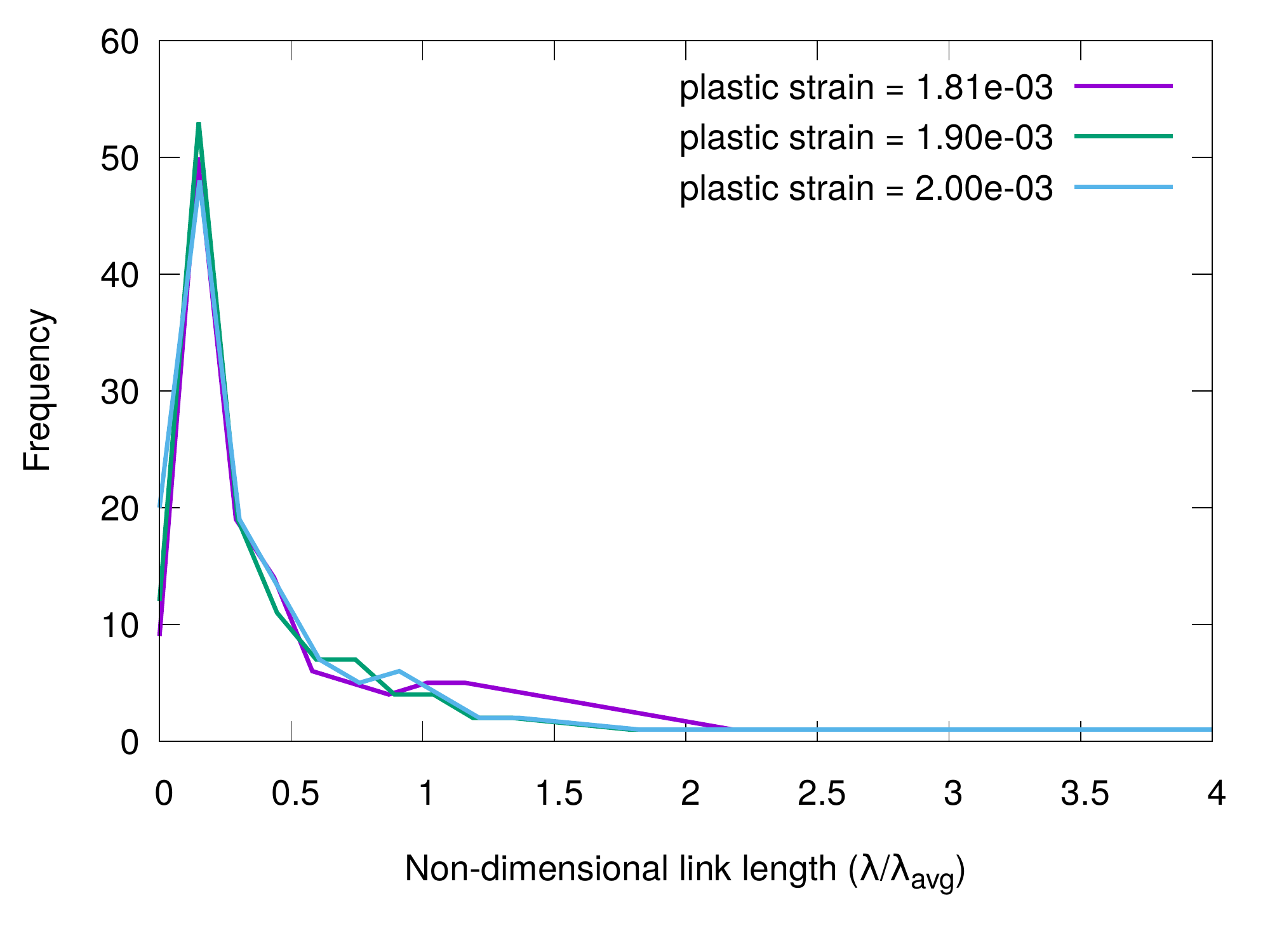}}
    \hspace{0mm}
    \subfloat[]{\includegraphics[width=7.0cm]{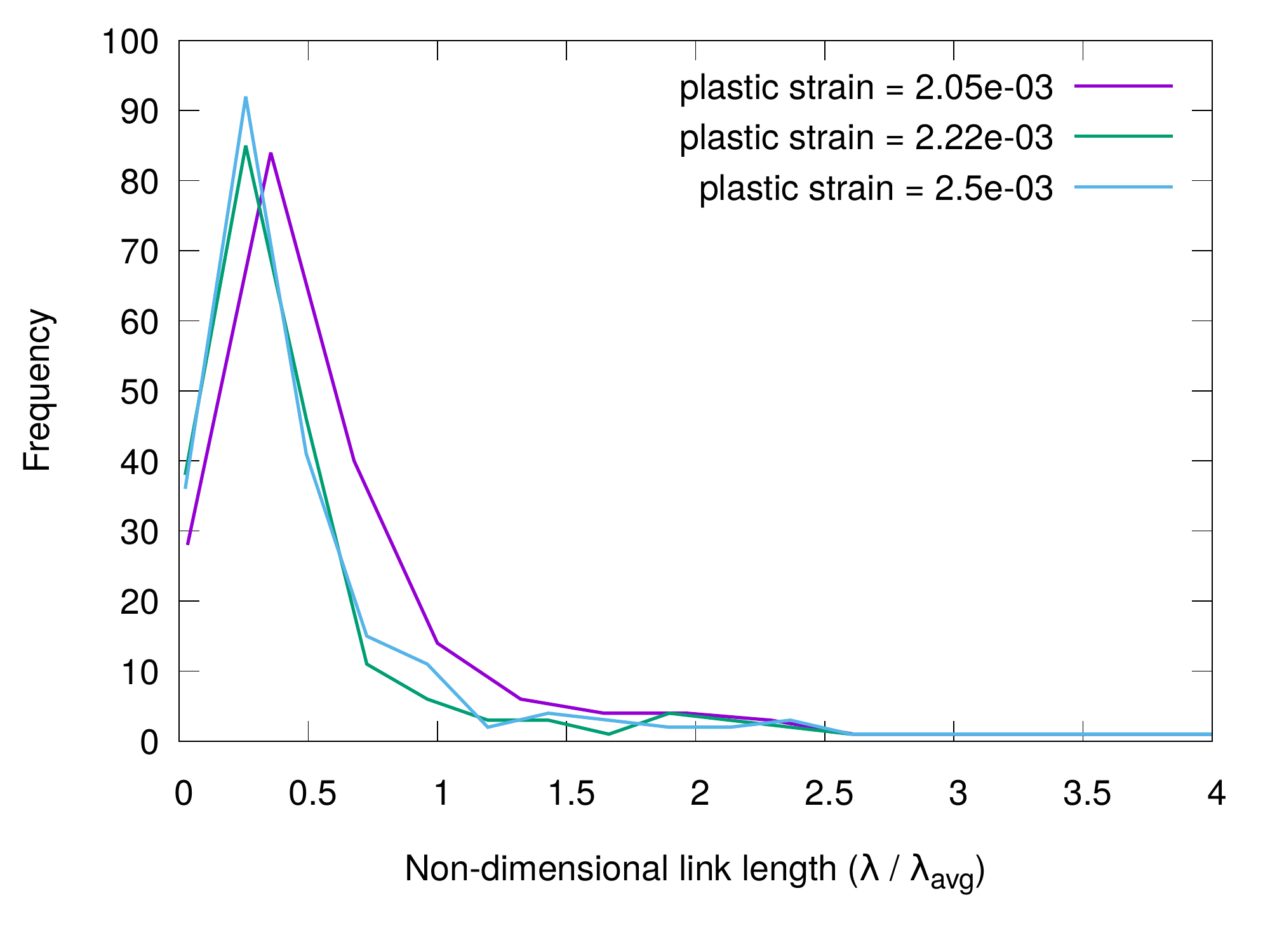}}
    \subfloat[]{\includegraphics[width=7.0cm]{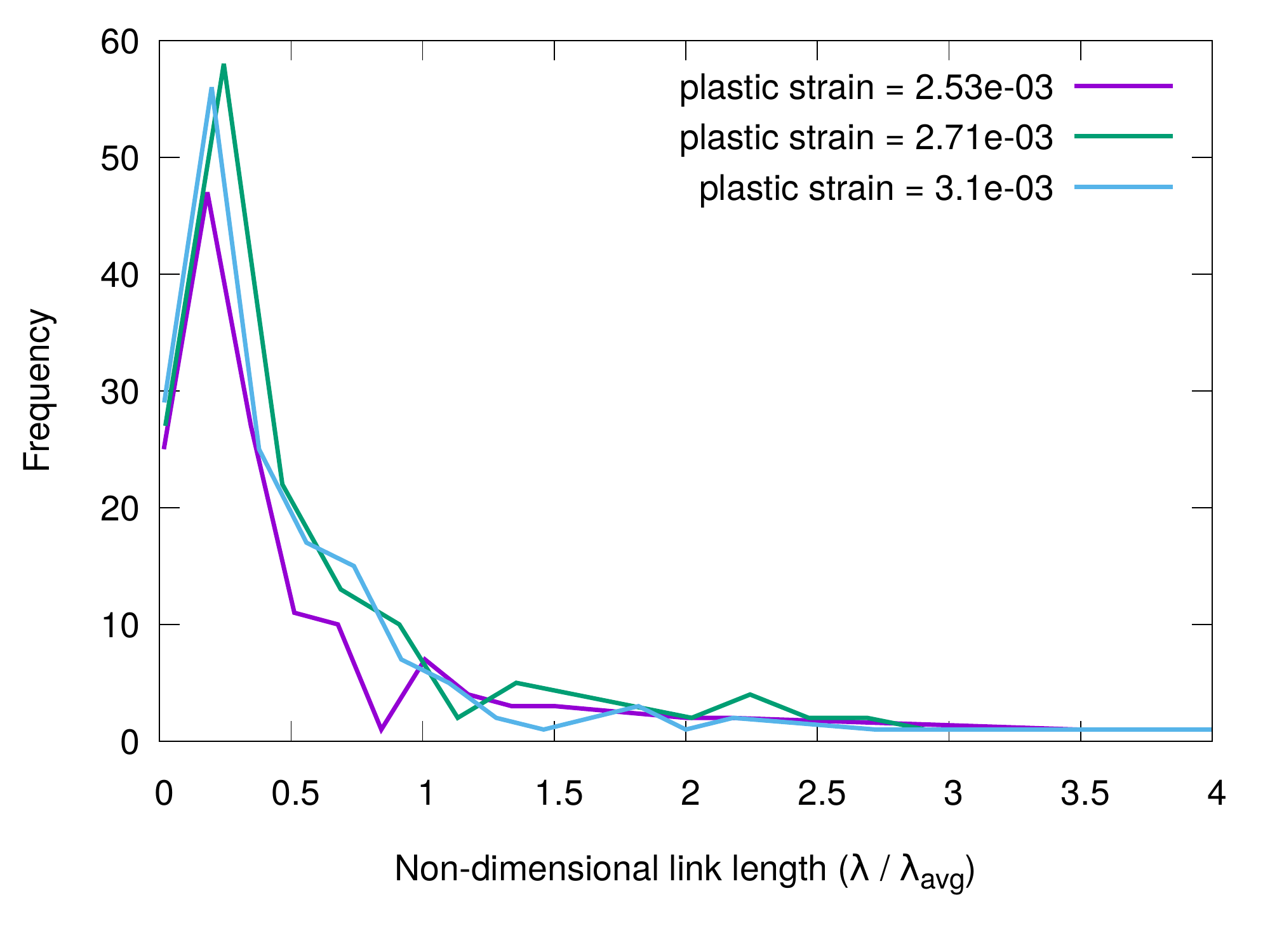}}
    \caption{
      (a) Evolution of ratio of mobile dislocation density to   
          immobile dislocation density ($\frac{\rho^m}{\rho^{im}}$) against 
          plastic strain. Steady state is reached when 
          $\frac{\rho^m}{\rho^{im}}$ tends to be constant. 
          Increase in applied stress delays the onset of steady state. 
          For $350 \; MPa$, steady state is reached close to strain level of 
          $0.00175$, whereas for $500\;MPa$, steady state is reached at 
          close to strain level of $0.0025$.
      (b) Immobile link length distribution for the case of 350 
          MPa. 
      (c) Immobile link length distribution for the case of 400
          MPa.
      (d) Immobile link length distribution for the case of 500    
          MPa             
    }
    \label{mobile_by_immobile}
\end{figure}

   
The IDN comprises mobile as 
well as immobile dislocations. The ratio of
mobile to immobile dislocation length ($\dfrac{\rho^m}{\rho^{im}}$) 
in the network is strongly influenced by applied stress, as
seen in fig \ref{mobile_by_immobile}(a). 
Figure \ref{mobile_by_immobile}(b),(c), and (d) show the distribution of 
immobile link length. It is evident that the immobile link length 
distribution is fairly time independent after a critical strain is achieved.
The immobile dislocation link length scales inversely with 
applied stress \cite{madec2000new}. The mobile dislocations can easily 
multiply in simulations with higher applied 
stresses. Consequently, larger mobile dislocation density has 
more probability to collide with network to form junctions, and thus 
the large number of triple junctions of shorter length are formed at 
higher stresses. 

\par
\begin{figure}[!htb]%
  \centering
  \includegraphics*[width=12cm]{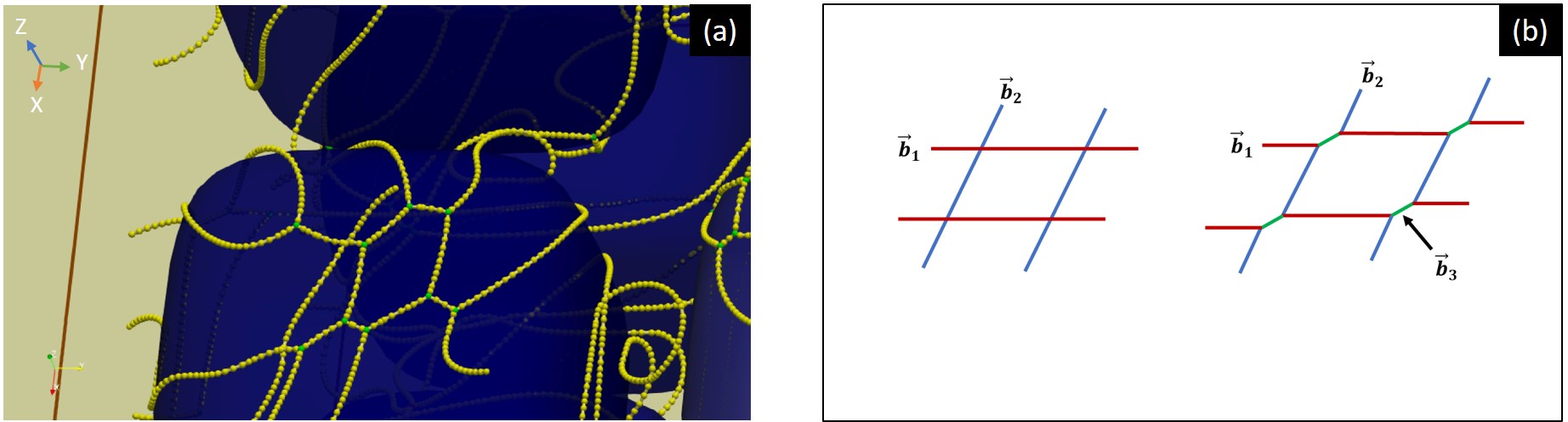}
  \caption{ 
      (a) Observation of hexagonal loops at stress level of 300 MPa.
          The loop is formed on face of precipitate oriented along
          $[001]$ crystallographic direction.
      (b) Schematic representation of formation of hexagonal loops. 
          When two parallel dislocation lines of Burgers vector 
          $\vec{b}_1$ and $\vec{b}_2$ intersects, the trinode 
          junction forms at intersection points. The junction formed 
          has burgers vector $\vec{b}_3$ which is different from 
          other two segments. When the dislocation segments of 
          $\vec{b}_3$ is larger then the loop appears to be of 
          hexagonal morphology whereas when it is shorter the 
          morphology of loop appears to be close to square.    
  }
  \label{hexaloop}
\end{figure}

Our simulation results show  
morphological features of interfacial dislocation 
network which bears good resemblance to that of the TEM observations.
We have observed hexagonal loops over the surface of
particles, comprised of dislocation segments with 
three different Burgers vector (refer fig \ref{hexaloop} (a) )
\cite{Gabb1989,Sugui2000}. 
The formation of hexagonal loops can be explained as illustrated in 
fig \ref{hexaloop} (b). In some instances, 
distorted hexagonal loops may form as an effect of immobile 
dislocation links being dragged due to motion of mobile 
dislocation links. Although, in case of 500 MPa, the 
loops consist of dislocation segments with 
three distinct Burgers vectors, the morphology
appears to be more square like. This can be attributed to 
the shorter length of immobile dislocation lengths at 
applied stress of 500 MPa.

\begin{figure}[!htb]%
 \centering
 \includegraphics*[width=12cm]{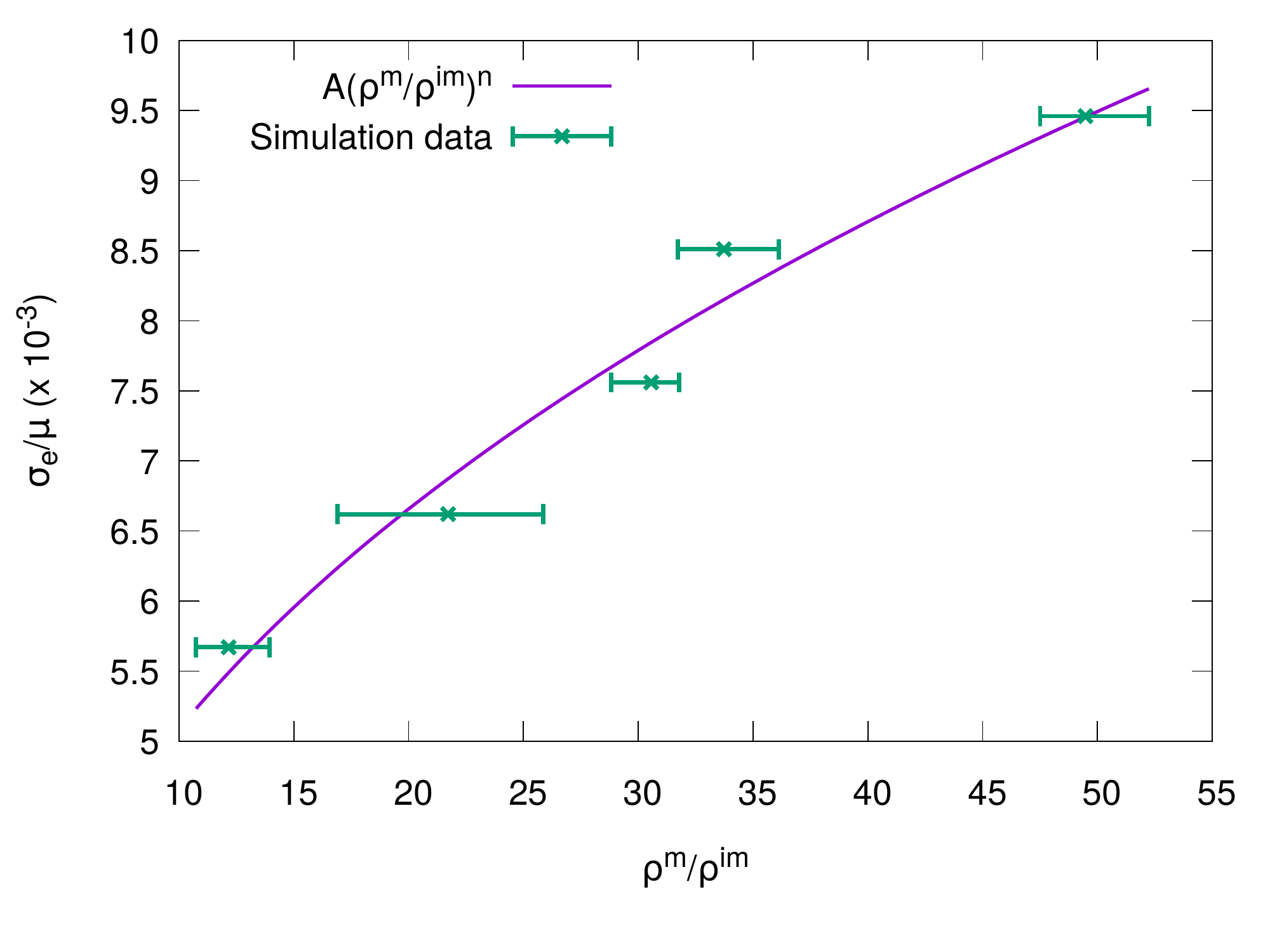}
 \caption{Effect of normalized applied stress on mobile to immobile
 dislocation density ratio. After fitting, the values of  
 $A = 2.087$ and $n = 0.3872$ are obtained. The graph shows that ratio of 
 mobile to immobile density scales non-linearly with applied stress.}
 \label{linear_relation}
\end{figure}%  
\par
Additionally, we show the relationship between the normalized effective 
stress ($\frac{\sigma_e}{\mu}$) and ratio of mobile to 
immobile dislocation density ($\frac{\rho^m}{\rho^{im}}$). The effective 
stress is calculated by subtracting volume average back stress from applied
stress. We invoke work hardening theory by Brown-Stobbs~\cite{Brown1971} 
to get an approximate measure of back-stress. The back-stress can be 
approximately given as product of shear modulus ($\mu$), 
plastic strain($\epsilon_p$), and volume fraction of particles. 
As per figure \ref{mobile_by_immobile}, after 
a critical strain is achieved, the $\frac{\rho^m}{\rho^{im}}$ reaches a steady
state values in all cases of different stress levels. 
Figure \ref{linear_relation} 
represents the effect of normalized applied stress on ratio of 
mobile to immobile dislocation density. The average values of 
$\frac{\rho^m}{\rho^{im}}$ after the steady state value is reached are 
obtained and fitted to equation \ref{fit}.

\begin{equation}
\left( \frac{\sigma_e}{\mu} \right) = A\left( \frac{\rho^m}{\rho^{im}} \right)^n
\label{fit}
\end{equation} 

After fitting, the values of $A$ and $n$ are obtained to be $2.087$
and $0.3872$ with standard error of $15.29\%$ and $11.39\%$ respectively.  
This result signifies that the normalized effective stress scales 
non-linearly with ratio of mobile to immobile dislocation density. 
This result displays the deviation from Taylor's relation where the exponent 
to dislocation density is half. The deviation from the Taylor's relation 
can be attributed to high temperature recovery mechanisms present in the 
model. 
In the light of our results, the exponent to dislocation density 
can be rationalized as per ref \cite{horiuchi1972mechanism} which 
explains the relationship between stress and dislocation density at high 
homologous temperature.
Assume that the rate of increase in dislocation density is proportional to 
plastic strain rate, i.e.,
\begin{equation}
\frac{\mathrm{d}\rho}{\mathrm{d}t } = \alpha \dot{\varepsilon}
\label{increment}
\end{equation}
where $\mathrm{d}\rho$ is change in dislocation density, $\dot{\varepsilon}$
is plastic strain, $\mathrm{d}t$ is time interval, and $\alpha$ is 
proportionality constant. Further, the annihilation of dislocation density at 
high homologous temperatures is dependent on dislocation spacing. The 
dislocation spacing $l_c$ can be approximated as 
$l_c \approx \frac{1}{\sqrt{\rho}}$. Hence, annihilation rate is given as
\begin{equation}
\frac{\mathrm{d}\rho}{\mathrm{d}t} = - \beta \frac{\rho}{l_c} = 
    - \beta \rho^{3/2}
\label{annihilation}
\end{equation}
where $\beta$ is proportionality constant. From equation \eqref{increment} 
and \eqref{annihilation}, we obtain the net rate of change of dislocation 
density as 
\begin{equation}
\frac{\mathrm{d}\rho}{\mathrm{d}t} = \alpha \dot{\varepsilon} - 
\beta \rho^{3/2}
\label{netrate}
\end{equation}
At steady state, where $\frac{\mathrm{d}\rho}{\mathrm{d}t} = 0$, from equation
equation \eqref{netrate}, we yield a relation between strain rate and 
dislocation density as
\begin{equation}
\dot{\varepsilon} = \left( \frac{\beta}{\alpha}\right) \rho^{3/2}
\label{strainrate_dislocationdensity}
\end{equation}
The stress exponent of single crystal superalloy CMSX-4 varies from 3 to 7
at low to intermediate stresses and high homologous temperatures 
\cite{kondo2007stress}. Hence
\begin{equation}
\dot{\varepsilon} = K \sigma_e^{3 - 7}
\label{creepequation}
\end{equation}
where $K$ is proportionality constant and $\sigma_e$ is effective stress.
Combining equations \eqref{strainrate_dislocationdensity} and 
\eqref{creepequation}, we obtain the relation between effecttive stress and
dislocation density as
\begin{equation}
\sigma_e =  \left( \frac{\beta}{\alpha K}\right)^{0.14-0.33} \rho^{0.21-0.5}
\label{stress_density_relation}
\end{equation}
The exponent we obtained in equation \eqref{fit} from simulation data
lies in between the exponent shown in equation 
\eqref{stress_density_relation}. 

To summarize, three dimensional DDD simulations are performed by 
employing the modified ParaDis code. Our simulations show the formation of 
three dimensional IDN around cuboidal particles. The IDN is 
formed as a by-product of dislocation reaction. The characteristics of
dislocation network morphology resembles well to the experiments. 
The attractive junction generated due to dislocation reactions provides the 
strength to the network. The immobile link length distribution is time independent
after a critical strain is achieved at given applied stress. Moreover, we present 
the relationship between applied stress and ratio of mobile to immobile 
dislocation density and it agrees well with theoretical relation. The relation shows
deviation from Taylor's relation.   

\section*{References}

\bibliography{mybibfile}

\begin{thebibliography}{38}
\expandafter\ifx\csname natexlab\endcsname\relax\def\natexlab#1{#1}\fi
\providecommand{\url}[1]{\texttt{#1}}
\providecommand{\href}[2]{#2}
\providecommand{\path}[1]{#1}
\providecommand{\DOIprefix}{doi:}
\providecommand{\ArXivprefix}{arXiv:}
\providecommand{\URLprefix}{URL: }
\providecommand{\Pubmedprefix}{pmid:}
\providecommand{\doi}[1]{\href{http://dx.doi.org/#1}{\path{#1}}}
\providecommand{\Pubmed}[1]{\href{pmid:#1}{\path{#1}}}
\providecommand{\bibinfo}[2]{#2}
\ifx\xfnm\relax \def\xfnm[#1]{\unskip,\space#1}\fi
%Type = Book
\bibitem[{Reed(2008)}]{reed2008superalloys}
\bibinfo{author}{R.~C. Reed}, \bibinfo{title}{The superalloys: fundamentals and
  applications}, \bibinfo{publisher}{Cambridge university press},
  \bibinfo{year}{2008}.
%Type = Article
\bibitem[{Lasalmonie and Strudel(1975)}]{Lasalmonie1975}
\bibinfo{author}{A.~Lasalmonie}, \bibinfo{author}{J.~L. Strudel},
  \bibinfo{journal}{Phil. Mag.} \bibinfo{volume}{32} (\bibinfo{year}{1975})
  \bibinfo{pages}{937--949}.
%Type = Article
\bibitem[{Feller-Kniepmeier and Link(1989)}]{Feller-Kniepmeier1989}
\bibinfo{author}{M.~Feller-Kniepmeier}, \bibinfo{author}{T.~Link},
  \bibinfo{journal}{Mat. Sci. and Engg. A} \bibinfo{volume}{113}
  (\bibinfo{year}{1989}) \bibinfo{pages}{191--195}.
%Type = Article
\bibitem[{Pollock and Argon(1992)}]{Pollock1992}
\bibinfo{author}{T.~Pollock}, \bibinfo{author}{A.~Argon},
  \bibinfo{journal}{Acta Metal. et Mat.} \bibinfo{volume}{40}
  (\bibinfo{year}{1992}) \bibinfo{pages}{1--30}.
%Type = Article
\bibitem[{Nategh and Sajjadi(2003)}]{Nategh2003}
\bibinfo{author}{S.~Nategh}, \bibinfo{author}{S.~A. Sajjadi},
  \bibinfo{journal}{Mat. Sci. and Engg. A} \bibinfo{volume}{339}
  (\bibinfo{year}{2003}) \bibinfo{pages}{103--108}.
%Type = Article
\bibitem[{Carroll et~al.(2008)Carroll, Feng, and
  Pollock}]{carroll2008interfacial}
\bibinfo{author}{L.~Carroll}, \bibinfo{author}{Q.~Feng},
  \bibinfo{author}{T.~Pollock}, \bibinfo{journal}{Metall. and Mat. Trans. A}
  \bibinfo{volume}{39} (\bibinfo{year}{2008}) \bibinfo{pages}{1290--1307}.
%Type = Article
\bibitem[{Ru et~al.(2016)Ru, Li, Zhou, Pei, Wang, Gong, and
  Xu}]{ru2016dislocation}
\bibinfo{author}{Y.~Ru}, \bibinfo{author}{S.~Li}, \bibinfo{author}{J.~Zhou},
  \bibinfo{author}{Y.~Pei}, \bibinfo{author}{H.~Wang},
  \bibinfo{author}{S.~Gong}, \bibinfo{author}{H.~Xu}, \bibinfo{journal}{Sci.
  rep.} \bibinfo{volume}{6} (\bibinfo{year}{2016}) \bibinfo{pages}{29941}.
%Type = Article
\bibitem[{Gabb et~al.(1989)Gabb, Draper, Hull, Mackay, and Nathal}]{Gabb1989}
\bibinfo{author}{T.~Gabb}, \bibinfo{author}{S.~Draper},
  \bibinfo{author}{D.~Hull}, \bibinfo{author}{R.~Mackay},
  \bibinfo{author}{M.~Nathal}, \bibinfo{journal}{Mat. Sci. and Engg.: A}
  \bibinfo{volume}{118} (\bibinfo{year}{1989}) \bibinfo{pages}{59--69}.
%Type = Article
\bibitem[{Sugui et~al.(2000)Sugui, Huihua, Jinghua, Hongcai, Yongbo, and
  Zhuangqi}]{Sugui2000}
\bibinfo{author}{T.~Sugui}, \bibinfo{author}{Z.~Huihua},
  \bibinfo{author}{Z.~Jinghua}, \bibinfo{author}{Y.~Hongcai},
  \bibinfo{author}{X.~Yongbo}, \bibinfo{author}{H.~Zhuangqi},
  \bibinfo{journal}{Mat. Sci. and Engg. A} \bibinfo{volume}{279}
  (\bibinfo{year}{2000}) \bibinfo{pages}{160--165}.
%Type = Article
\bibitem[{Zhang et~al.(2002)Zhang, Murakumo, Koizumi, Kobayashi, Harada, and
  Masaki}]{Zhang2002}
\bibinfo{author}{J.~X. Zhang}, \bibinfo{author}{T.~Murakumo},
  \bibinfo{author}{Y.~Koizumi}, \bibinfo{author}{T.~Kobayashi},
  \bibinfo{author}{H.~Harada}, \bibinfo{author}{S.~Masaki},
  \bibinfo{journal}{Metall. and Mat. Tran. A: Phys. Metall. and Mat. Sci.}
  \bibinfo{volume}{33} (\bibinfo{year}{2002}) \bibinfo{pages}{3741--3746}.
%Type = Article
\bibitem[{Zhang et~al.(2003)Zhang, Murakumo, Harada, and Koizumi}]{Zhang2003}
\bibinfo{author}{J.~X. Zhang}, \bibinfo{author}{T.~Murakumo},
  \bibinfo{author}{H.~Harada}, \bibinfo{author}{Y.~Koizumi},
  \bibinfo{journal}{Scr. Mat.} \bibinfo{volume}{48} (\bibinfo{year}{2003})
  \bibinfo{pages}{287--293}.
%Type = Article
\bibitem[{Zhang et~al.(2005)Zhang, Wang, Harada, and Koizumi}]{Zhang2005}
\bibinfo{author}{J.~X. Zhang}, \bibinfo{author}{J.~C. Wang},
  \bibinfo{author}{H.~Harada}, \bibinfo{author}{Y.~Koizumi},
  \bibinfo{journal}{Acta Mat.} \bibinfo{volume}{53} (\bibinfo{year}{2005})
  \bibinfo{pages}{4623--4633}.
%Type = Article
\bibitem[{Alexander et~al.(2007)Alexander, Thomas, and Gert}]{Epishin2007}
\bibinfo{author}{E.~Alexander}, \bibinfo{author}{L.~Thomas},
  \bibinfo{author}{N.~Gert}, \bibinfo{journal}{J. of Micro.}
  \bibinfo{volume}{228} (\bibinfo{year}{2007}) \bibinfo{pages}{110--117}.
%Type = Article
\bibitem[{Lagneborg(1968)}]{Lagneborg1968}
\bibinfo{author}{R.~Lagneborg}, \bibinfo{journal}{J. of Mat. Sci.}
  \bibinfo{volume}{3} (\bibinfo{year}{1968}) \bibinfo{pages}{596--602}.
%Type = Article
\bibitem[{Lagneborg(1969{\natexlab{a}})}]{Lagneborg1969}
\bibinfo{author}{R.~Lagneborg}, \bibinfo{journal}{Metal Sci. J.}
  \bibinfo{volume}{3} (\bibinfo{year}{1969}{\natexlab{a}})
  \bibinfo{pages}{18--23}.
%Type = Article
\bibitem[{Lagneborg(1969{\natexlab{b}})}]{Lagneborg1969a}
\bibinfo{author}{R.~Lagneborg}, \bibinfo{journal}{Metal Sci. J.}
  \bibinfo{volume}{3} (\bibinfo{year}{1969}{\natexlab{b}})
  \bibinfo{pages}{161--168}.
%Type = Article
\bibitem[{Yue et~al.(2018)Yue, Liu, Yang, Huang, Zhang, and Fu}]{yue2018stress}
\bibinfo{author}{Q.~Yue}, \bibinfo{author}{L.~Liu}, \bibinfo{author}{W.~Yang},
  \bibinfo{author}{T.~Huang}, \bibinfo{author}{J.~Zhang},
  \bibinfo{author}{H.~Fu}, \bibinfo{journal}{Mat. Sci. and Engg. A}
  (\bibinfo{year}{2018}).
%Type = Article
\bibitem[{Field et~al.(1992)Field, Pollockf, and Murphy}]{field1992development}
\bibinfo{author}{R.~Field}, \bibinfo{author}{T.~Pollockf},
  \bibinfo{author}{W.~Murphy}, \bibinfo{journal}{Superalloys 1992}
  (\bibinfo{year}{1992}).
%Type = Article
\bibitem[{Yashiro et~al.(2008)Yashiro, Konishi, and Tomita}]{Yashiro2008}
\bibinfo{author}{K.~Yashiro}, \bibinfo{author}{M.~Konishi},
  \bibinfo{author}{Y.~Tomita}, \bibinfo{journal}{Comp. Mat. Sci.}
  \bibinfo{volume}{43} (\bibinfo{year}{2008}) \bibinfo{pages}{481--488}.
%Type = Article
\bibitem[{Huang et~al.(2012)Huang, Zhao, and Tong}]{Huang2012}
\bibinfo{author}{M.~Huang}, \bibinfo{author}{L.~Zhao},
  \bibinfo{author}{J.~Tong}, \bibinfo{journal}{Intl. J. of Plast.}
  \bibinfo{volume}{28} (\bibinfo{year}{2012}) \bibinfo{pages}{141--158}.
%Type = Article
\bibitem[{{Hafez Haghighat} et~al.(2013){Hafez Haghighat}, Eggeler, and
  Raabe}]{HafezHaghighat2013}
\bibinfo{author}{S.~M. {Hafez Haghighat}}, \bibinfo{author}{G.~Eggeler},
  \bibinfo{author}{D.~Raabe}, \bibinfo{journal}{Acta Mat.} \bibinfo{volume}{61}
  (\bibinfo{year}{2013}) \bibinfo{pages}{3709--3723}.
%Type = Article
\bibitem[{Liu et~al.(2014)Liu, Raabe, Roters, and Arsenlis}]{Liu2014}
\bibinfo{author}{B.~Liu}, \bibinfo{author}{D.~Raabe},
  \bibinfo{author}{F.~Roters}, \bibinfo{author}{A.~Arsenlis},
  \bibinfo{journal}{Acta Mat.} \bibinfo{volume}{79} (\bibinfo{year}{2014})
  \bibinfo{pages}{216--233}.
%Type = Article
\bibitem[{Gao et~al.(2015)Gao, Fivel, Ma, and Hartmaier}]{gao2015influence}
\bibinfo{author}{S.~Gao}, \bibinfo{author}{M.~Fivel}, \bibinfo{author}{A.~Ma},
  \bibinfo{author}{A.~Hartmaier}, \bibinfo{journal}{J. of the Mech. and Phy. of
  Solids} \bibinfo{volume}{76} (\bibinfo{year}{2015})
  \bibinfo{pages}{276--290}.
%Type = Article
\bibitem[{Zhu et~al.(2013)Zhu, Li, and Huang}]{zhu2013atomistic}
\bibinfo{author}{Y.~Zhu}, \bibinfo{author}{Z.~Li}, \bibinfo{author}{M.~Huang},
  \bibinfo{journal}{Comp. Mat. Sci.} \bibinfo{volume}{70}
  (\bibinfo{year}{2013}) \bibinfo{pages}{178--186}.
%Type = Article
\bibitem[{Lin et~al.(1989)Lin, Lee, and Ardell}]{lin1989scaling}
\bibinfo{author}{P.~Lin}, \bibinfo{author}{S.~Lee},
  \bibinfo{author}{A.~Ardell}, \bibinfo{journal}{Acta Metall.}
  \bibinfo{volume}{37} (\bibinfo{year}{1989}) \bibinfo{pages}{739--748}.
%Type = Article
\bibitem[{Shi and Northwood(1993)}]{shi1993dislocation}
\bibinfo{author}{L.~Shi}, \bibinfo{author}{D.~Northwood}, \bibinfo{journal}{J.
  of Mat. Sci.} \bibinfo{volume}{28} (\bibinfo{year}{1993})
  \bibinfo{pages}{5963--5974}.
%Type = Article
\bibitem[{Ardell and Przystupa(1984)}]{ardell1984dislocation}
\bibinfo{author}{A.~Ardell}, \bibinfo{author}{M.~Przystupa},
  \bibinfo{journal}{Mech. of Mat.} \bibinfo{volume}{3} (\bibinfo{year}{1984})
  \bibinfo{pages}{319--332}.
%Type = Article
\bibitem[{Ostrom and Lagneborg(1976)}]{ostrom1976recovery}
\bibinfo{author}{P.~Ostrom}, \bibinfo{author}{R.~Lagneborg},
  \bibinfo{journal}{J. of Engg. Mat. and Tech.} \bibinfo{volume}{98}
  (\bibinfo{year}{1976}) \bibinfo{pages}{114--121}.
%Type = Article
\bibitem[{Sills et~al.(2018)Sills, Bertin, Aghaei, and Cai}]{Sills2017}
\bibinfo{author}{R.~B. Sills}, \bibinfo{author}{N.~Bertin},
  \bibinfo{author}{A.~Aghaei}, \bibinfo{author}{W.~Cai},
  \bibinfo{journal}{Phys. Rev. Lett.} \bibinfo{volume}{121}
  (\bibinfo{year}{2018}) \bibinfo{pages}{085501}.
%Type = Article
\bibitem[{Mecking and Kocks(1981)}]{mecking1981kinetics}
\bibinfo{author}{H.~Mecking}, \bibinfo{author}{U.~Kocks},
  \bibinfo{journal}{Acta Metall.} \bibinfo{volume}{29} (\bibinfo{year}{1981})
  \bibinfo{pages}{1865--1875}.
%Type = Book
\bibitem[{Bulatov and Cai(2006)}]{bulatov2006computer}
\bibinfo{author}{V.~Bulatov}, \bibinfo{author}{W.~Cai},
  \bibinfo{title}{Computer simulations of dislocations},
  volume~\bibinfo{volume}{3}, \bibinfo{publisher}{Oxford University Press on
  Demand}, \bibinfo{year}{2006}.
%Type = Article
\bibitem[{Arsenlis et~al.(2007)Arsenlis, Cai, Tang, Rhee, Oppelstrup, Hommes,
  Pierce, and Bulatov}]{Arsenlis2007}
\bibinfo{author}{A.~Arsenlis}, \bibinfo{author}{W.~Cai},
  \bibinfo{author}{M.~Tang}, \bibinfo{author}{M.~Rhee},
  \bibinfo{author}{T.~Oppelstrup}, \bibinfo{author}{G.~Hommes},
  \bibinfo{author}{T.~G. Pierce}, \bibinfo{author}{V.~V. Bulatov},
  \bibinfo{journal}{Model. and Sim. in Mat. Sci. and Engg.}
  \bibinfo{volume}{15} (\bibinfo{year}{2007}) \bibinfo{pages}{553--595}.
%Type = Article
\bibitem[{Jogi and Bhattacharya(2016)}]{jogi2016evolution}
\bibinfo{author}{T.~Jogi}, \bibinfo{author}{S.~Bhattacharya},
  \bibinfo{journal}{Trans. of the Indian Inst. of Metals} \bibinfo{volume}{69}
  (\bibinfo{year}{2016}) \bibinfo{pages}{507--512}.
%Type = Book
\bibitem[{Jaklic et~al.(2013)Jaklic, Leonardis, and
  Solina}]{jaklic2013segmentation}
\bibinfo{author}{A.~Jaklic}, \bibinfo{author}{A.~Leonardis},
  \bibinfo{author}{F.~Solina}, \bibinfo{title}{Segmentation and recovery of
  superquadrics}, volume~\bibinfo{volume}{20}, \bibinfo{publisher}{Springer
  Science \& Business Media}, \bibinfo{year}{2013}.
%Type = Article
\bibitem[{Madec et~al.(2000)Madec, Devincre, and Kubin}]{madec2000new}
\bibinfo{author}{R.~Madec}, \bibinfo{author}{B.~Devincre},
  \bibinfo{author}{L.~P. Kubin}, \bibinfo{journal}{MRS Online Proc. Lib. Arc.}
  \bibinfo{volume}{653} (\bibinfo{year}{2000}).
%Type = Article
\bibitem[{Brown and Stobbs(1971)}]{Brown1971}
\bibinfo{author}{L.~M. Brown}, \bibinfo{author}{W.~M. Stobbs},
  \bibinfo{journal}{Phil. Mag}  (\bibinfo{year}{1971})
  \bibinfo{pages}{1185--1199}.
%Type = Article
\bibitem[{Horiuchi and Otsuka(1972)}]{horiuchi1972mechanism}
\bibinfo{author}{R.~Horiuchi}, \bibinfo{author}{M.~Otsuka},
  \bibinfo{journal}{Trans. of the Japan Inst. of Metals} \bibinfo{volume}{13}
  (\bibinfo{year}{1972}) \bibinfo{pages}{284--293}.
%Type = Inproceedings
\bibitem[{Kondo et~al.(2007)Kondo, Miura, and Matsuo}]{kondo2007stress}
\bibinfo{author}{Y.~Kondo}, \bibinfo{author}{N.~Miura},
  \bibinfo{author}{T.~Matsuo}, in: \bibinfo{booktitle}{Mat. Sci. Forum}, volume
  \bibinfo{volume}{539}, \bibinfo{organization}{Trans Tech Publ}, pp.
  \bibinfo{pages}{3100--3105}.

\end{thebibliography}
%\bibliography{short,cvpubs}
\end{document}